\documentclass[journal]{IEEEtran}
\usepackage{cite}

\ifCLASSINFOpdf 
\usepackage[pdftex]{graphicx}
\else
\usepackage[dvips]{graphicx}
\fi

\usepackage[cmex10]{amsmath}
\usepackage{algorithmic}
\usepackage[ruled,norelsize]{algorithm2e}

\usepackage{array}

\usepackage{mdwmath}
\usepackage{mdwtab}
\usepackage{todonotes}

\usepackage{eqparbox}

\ifCLASSOPTIONcompsoc
\usepackage[caption=false,font=normalsize,labelfont=sf,textfont=sf]{subfig}
\else
\usepackage[caption=false,font=footnotesize]{subfig}
\fi

\usepackage{fixltx2e}
\usepackage{url}
\usepackage{graphicx}
\usepackage{amsfonts}
\usepackage{amssymb}
\usepackage{amsthm}
\usepackage{bm}
\usepackage[ruled]{algorithm2e}
\usepackage[utf8]{inputenc}
\usepackage{booktabs}
\usepackage{epstopdf}
\usepackage{tabularx}
\usepackage{cite}
\usepackage{flushend}
\usepackage{multirow}
\usepackage{xcolor,colortbl}

\definecolor{maroon}{cmyk}{0,0.87,0.68,0.32}

\newcounter{tempEquationCounter} 
\newcounter{thisEquationNumber}

\begin{document}
%
\title{Exploiting CSMA/ECA and Adaptive Sensitivity Control for Simultaneous Transmit and Receive\\ in IEEE 802.11 WLANs}

\author{Adnan~Aijaz\(^\dagger\) and Parag~Kulkarni\(^\ddagger\)  \\
\(^\dagger\)Toshiba Research Europe Ltd., Bristol, U.K. \quad \(^\ddagger\)United Arab Emirates University, Al-Ain, Abu Dhabi, U.A.E. \\
\thanks{This work was done while P. Kulkarni was working at Toshiba Research Europe Ltd., U.K.}%
}

\maketitle
\begin{abstract}
\boldmath
Ever since the feasibility of in-band full-duplex (FD)  at the Physical (PHY) layer has been established, several studies have emerged investigating protocol aspects of enabling FD operation in various legacy wireless technologies. 
 Recently, the adoption of  a simultaneous transmit and receive (STR) mode for next generation wireless local area networks (WLANs) has received significant attention. With STR mode (FD communication mode) in 802.11 WLANs,  bi-directional FD (BFD) and uni-directional FD (UFD) links are created. STR mode in 802.11 WLANs must be enabled with minimal protocol modifications while accounting for the co-existence and compatibility with legacy nodes and protocols. This paper provides a novel solution, that can leverage carrier sense multiple access with enhanced collision avoidance (CSMA/ECA) and adaptive sensitivity control mechanisms, for enabling STR operation. The key aspects of the proposed solution include co-existence with legacy nodes, identification of eligible nodes for UFD, optimization of secondary BFD and UFD transmissions, and creation of UFD opportunities. Performance evaluation demonstrates that the proposed solution is effective in achieving the gains provided by STR operation.


\end{abstract}


\vspace{-9pt}

%
\IEEEpeerreviewmaketitle

\section{Introduction}
\IEEEPARstart{T}{raditional}  IEEE 802.11 wireless local area networks (WLANs), which employ the carrier sense multiple access with collision avoidance (CSMA/CA) protocol, continue to be a popular choice for providing cost-effective wireless Internet access. Some of the challenges faced by the next generation WLANs \cite{survey_hew} include supporting  the explosive growth in wireless Internet traffic, providing increased user throughputs and successful operation in  dense deployments. This necessitates  solutions to enhance efficiency in increasingly challenging environments. Motivated by this,  there have been efforts along various dimensions, e.g., improving the efficiency of channel access through carrier sense multiple access with enhanced collision avoidance (CSMA/ECA), enhancing spatial reuse opportunities through sensitivity adaptation, and more recently, simultaneous transmit and receive (STR) operation for maximizing medium utilization and throughput. 

\begin{figure}
\centering
\includegraphics[scale=0.26]{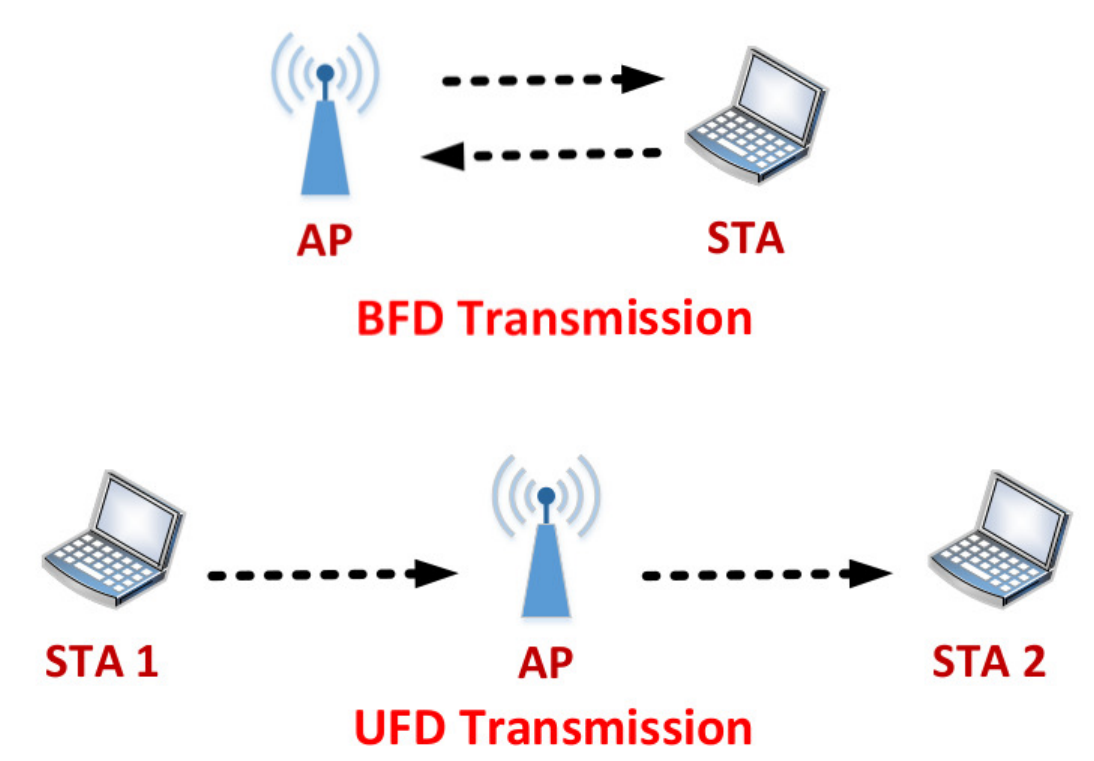}
\caption{An illustration of STR mode in 802.11 networks.}
\label{fd_scenarios}
\end{figure}

Recent developments in self-interference cancellation techniques   have made in-band full-duplex (FD)  \cite{FD_SIC} operation feasible for wireless communications. Unlike half-duplex (HD) radios, \textcolor{black}{FD radios are capable of simultaneous transmission and reception on the same frequency resource. Recent studies on FD wireless networks have mainly focused on Physical (PHY) layer aspects; however,  medium access control (MAC) and network layer protocols  have also started to emerge \cite{FD_Phy_Mac}, \cite{xfdr_wcm}. }
The capability of FD operation in legacy wireless  technologies like WLANs provides a number of benefits. 
The STR mode theoretically doubles the channel capacity of WLANs.  STR operation in WLANs creates two distinct types of wireless links (illustrated in Fig. \ref{fd_scenarios}): (a) bi-directional FD (BFD) in which a pair of FD-capable access point (AP) and station (STA) can simultaneously transmit/receive to/from each other, \textcolor{black}{(b) unidirectional FD (UFD) wherein the AP can simultaneously engage in transmission with a FD/HD STA while receiving from another FD/HD STA.}



The provisioning of STR mode in 802.11 WLANs creates various challenges. These challenges, which have been discussed in detail in our previous work \cite{aijaz_str_wcm}, include co-existence of FD and HD nodes (APs and STAs), enabling both BFD and UFD transmissions without modifications to legacy channel access procedures, selecting eligible nodes for UFD transmission, acknowledgement (ACK) timeout setting for nodes engaged in simultaneous transmission and reception, and contention unfairness for nodes overhearing FD transmissions. In \cite{aijaz_str_wcm}, we designed a simple and practical solution to address these challenges while accounting for FD/HD co-existence and backward compatibility.

This paper proposes a novel solution that exploits CSMA/ECA and adaptive sensitivity control mechanisms to achieve STR operation in 802.11 WLANs. \textcolor{black}{To the best of our knowledge, this is the first study that brings together these (rather unrelated) techniques in context of enabling STR mode for 802.11 WLANs. The proposed solution is termed as \textsf{MASTER}\footnote{\underline{M}edium \underline{A}ccess control for \underline{S}imultaneous \underline{T}ransmit and receive with \underline{E}nhanced collision avoidance and sensitivity cont\underline{r}ol.}.  Unlike our previous solution  \cite{aijaz_str_wcm} and various related studies, \textsf{MASTER} does not rely on a request-to-send/clear-to-send (RTS/CTS) handshake mechanism for initiating FD transmissions. Other distinguishing aspects of \textsf{MASTER} include co-existence with legacy HD as well as legacy CSMA/CA-capable nodes, a standards compatible method for identifying eligible nodes for UFD scenario, the optimization of secondary transmission as part of BFD and UFD transmissions, and most notably the creation of UFD opportunities.} We begin our discussion by covering the  preliminaries and the state-of-the-art. This is followed by the proposed design framework for \textsf{MASTER}. Subsequently, we conduct a performance evaluation of the proposed protocol. Finally, the paper is concluded with a discussion on key insights.

\section{Background}\label{sect_chal}
 
\subsection{Legacy Protocol Operation}
\textcolor{black}{The legacy  WLAN operation is based on  IEEE 802.11 distributed coordination function (DCF) specification \cite{DCF} which adopts carrier sense multiple access with collision avoidance (CSMA/CA) and a binary exponential backoff procedure. } The AP periodically broadcasts the beacon frames containing network information. Typically, a STA associates with the AP through exchange of association request and response frames. Before initiating data transmission, a STA senses the medium and transmits only if it is idle. In case of a collision,  the transmitting STA takes a random backoff, which is uniformly distributed in the interval, \(\left[0, 2^k CW_{\text{min}}-1 \right]\) i.e., \(\mathcal{B}_{r} \leftarrow \mathcal{U}\left[0, 2^k CW_{\text{min}}-1 \right]\), where \(k\) denotes the backoff stage and \(CW_{\text{min}}\) is the minimum contention window. In case of a successful transmission, the receiving STA generates an ACK after waiting for short interframe space (SIFS) duration. After a successful transmission,  the transmitter and the receiver STAs wait for DCF interframe space (DIFS) duration before next contention.   
 
\subsection{CSMA/ECA} \label{info_eca}
The conventional CSMA/CA-based channel access is prone to collisions, and therefore, it  suffers from throughput degradation, especially in dense deployments. The IEEE 802.11ax working group has been considering modifications to CSMA/CA for achieving high throughput under dense deployments. A suitable alternative is the CSMA/ECA scheme which makes a simple modification to the legacy contention mechanism, thereby maintaining backward compatibility. Unlike CSMA/CA, STAs employ a deterministic backoff after \emph{successful} transmissions which is given by \(\mathcal{B}_{d} \leftarrow   \lceil CW_{\text{min}}/2 \rceil - 1 \). \textcolor{black}{The use of a deterministic backoff results in construction of a collision-free schedule among successful contenders in a distributed manner. It ensures that more channel time is spent in successful transmissions rather than recovering from collisions which leads to CSMA/ECA achieving higher throughput than CSMA/CA. Note that CSMA/ECA still uses the random backoff in case of collisions. Recently, a number of studies have investigated modeling and analysis of  CSMA/ECA. Our objective in this paper is to adopt CSMA/ECA as a tool for enabling STR operation. Therefore, the analytical aspects of CSMA/ECA are beyond the scope of this paper. A comprehensive performance evaluation of CSMA/ECA under non-saturated traffic and channel errors, and in co-existence with CSMA/CA nodes has been conducted in \cite{CSMA_ECA_Perf1}.  }

\subsection{Adaptive Sensitivity Control}
The IEEE 802.11-based radios utilize a clear channel assessment (CCA) module to sense the state of the medium. The IEEE 802.11ax working group has been actively investigating dynamic CCA modifications through dynamic (adaptive) sensitivity control (DSC) techniques. The fundamental principle of such sensitivity control techniques is to dynamically tune the carrier sensing threshold (CST). The adaptation of CST results in the co-existence of multiple concurrent transmitters, thereby improving spatial reuse \cite{aijaz_DSC}. While DSC techniques enhance spatial reuse in overlapping basic service sets (OBSSs), the tuning of CST in an adaptive manner is particularly attractive for STR operation as explained later.    An in-depth survey of adaptive carrier sensing techniques has been conducted in \cite{ACS_Survey}. Similarly, a comprehensive survey of various DSC techniques for improving spatial reuse in next generation WLANs has been conducted in \cite{survey_hew}.



\subsection{FD Protocols for WLANs: State-of-the-Art}\label{sect_rw}
In addition to our recent work \cite{aijaz_str_wcm}, various other studies have investigated MAC protocols for enabling FD communications in IEEE 802.11 WLANs. Tang and Wang proposed A-Duplex \cite{a-duplex}, \textcolor{black}{which has been designed for the co-existence of HD STAs and an FD AP}. Hence, it only supports UFD transmissions that are enabled through a handshake mechanism. A-Duplex relies on packet-alignment based capture effect for establishment of dual-links between the AP and two different STAs. It requires building an interference map of the network which may not be realized through standard compatibility. Besides, it introduces a new field in control frame header which affects backward compatibility.  Duarte \emph{et al.} \cite{fd_mac} proposed a MAC protocol for co-existence of HD and FD nodes. The protocol also requires an RTS/CTS handshake mechanism for FD discovery and opportunistic data transmissions. It only supports BFD transmissions. However, it is  not completely backward compatible due to modifications to ACK management and overhearing behavior of legacy STAs.  Choi \emph{et al.} \cite{PoCMAC} developed a power-controlled MAC (PoC-MAC) wherein only the AP operates in FD mode. Therefore,  it only supports UFD transmissions. PoC-MAC is not completely backward compatible as it introduces a signal strength based backoff mechanism and additional control frames for coordinating and completing FD transmissions. In \cite{scw-fd}, Marlali and Gurbuz proposed a synchronized contention window FD (S-CW FD) MAC protocol. S-CW FD MAC  accounts for the co-existence of FD and HD STAs. However, it requires certain modifications, including exchange of backoff window size information, that affect backward compatibility.


In summary, existing FD protocols for WLANs protocols are either not
backward/standard compatible or require handshake mechanisms for enabling FD operation.
 Further none of the existing proposals exploit the efficiency gains offered by CSMA/ECA. This paper aims to fill these gaps. 
 



\section{\textsf{MASTER} -- Key Aspects}\label{sect_protocol}
\textsf{MASTER} has a number of distinguishing features which are  summarized as follows. 

\begin{itemize}
\item \textbf{Co-existence} -- \textsf{MASTER} not only accounts for the co-existence of FD and legacy HD STAs in the network but also for the co-existence of nodes supporting CSMA/ECA and conventional CSMA/CA mechanisms. 

\item \textbf{Handshake-free STR} --  Unlike  existing studies (e.g., \cite{a-duplex}, \cite{fd_mac}, \cite{PoCMAC}, \cite{scw-fd}, \cite{aijaz_str_wcm}, and \cite{FD_WCL}), \textsf{MASTER} does not require an RTS/CTS handshake mechanism to initiate FD transmissions. It exploits the key capability of CSMA/ECA, i.e., collision-free schedule build-up, for achieving STR operation. \textcolor{black}{The handshake-free STR operation is crucial as the use of RTS/CTS results in degradation of application layer throughput. Besides, RTS/CTS frames are typically disabled in legacy WLANs.}

\item \textbf{Identification of Eligible Nodes} -- It is particularly important for the AP to know which STAs are eligible to become part of a UFD transmission. \textsf{MASTER} introduces a standards compatible method for the AP to acquire this information. 

\item \textbf{Secondary Transmission Optimization} -- \textcolor{black}{While achieving STR operation, it is particularly important to utilize  maximum possible duration for the secondary\footnote{\textcolor{black}{Any FD transmission (BFD or UFD) consists of two transmissions. The first and second transmissions are commonly referred to as primary and secondary transmissions, respectively.}} transmission.} Any processing at the AP, while a primary transmission is going on, may result in reduced duration for the secondary transmission. \textsf{MASTER} implements a simple and novel mechanism that ensures maximum possible duration for the secondary transmission.

\item \textbf{Creation of UFD Opportunities} -- \textsf{MASTER} implements a novel mechanism, based on the principle of adaptive sensitivity control, that can potentially create opportunities for UFD transmissions in the network. By exploiting such UFD opportunities, the achievable gain of STR can be maximized. 

\end{itemize}

\begin{figure*}
\label{STR_T}
\centering
\subfloat[]{\label{STR_txn1}\includegraphics[scale=0.28]{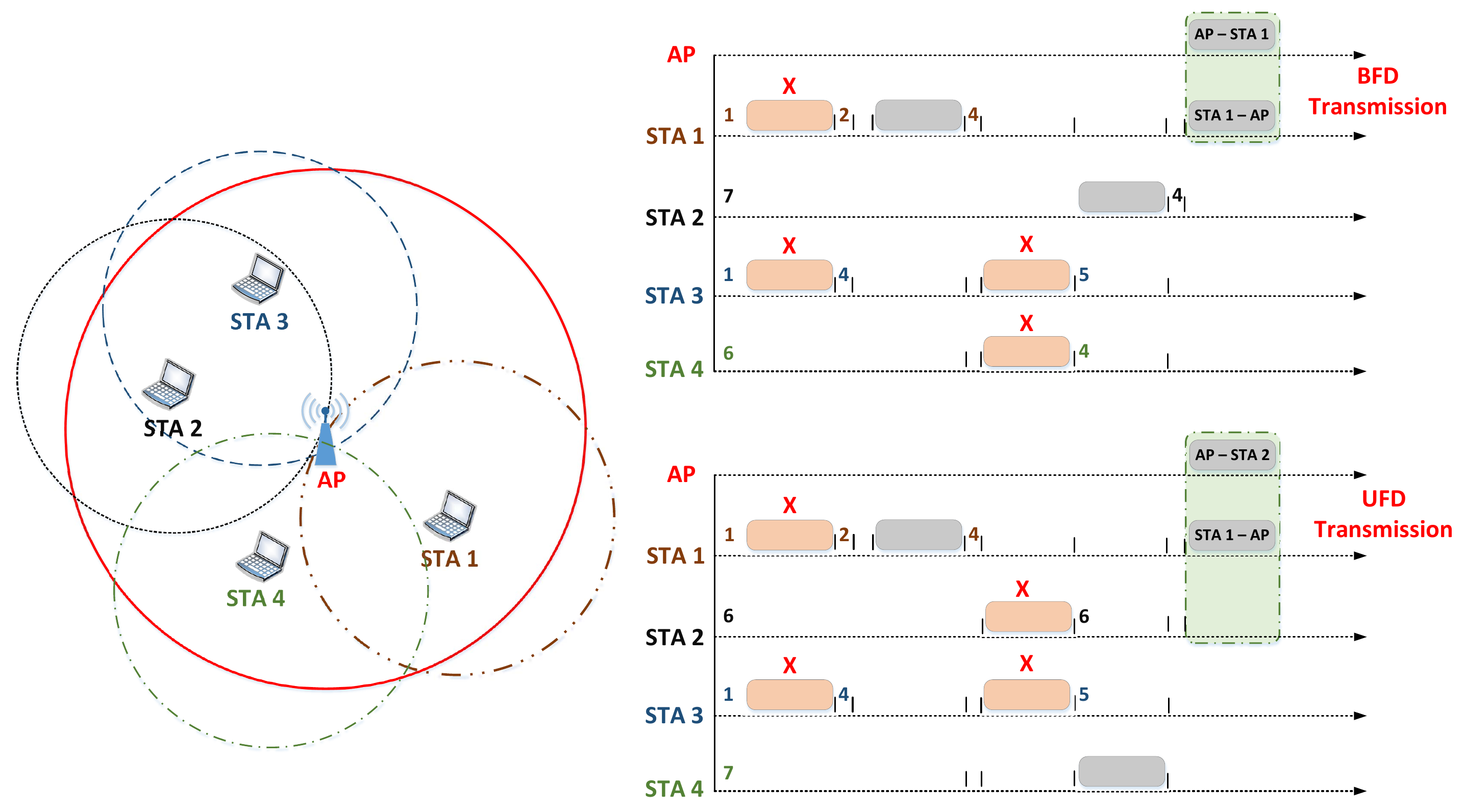}}  \quad
\subfloat[]{\label{STR_txn2}\includegraphics[scale=0.28]{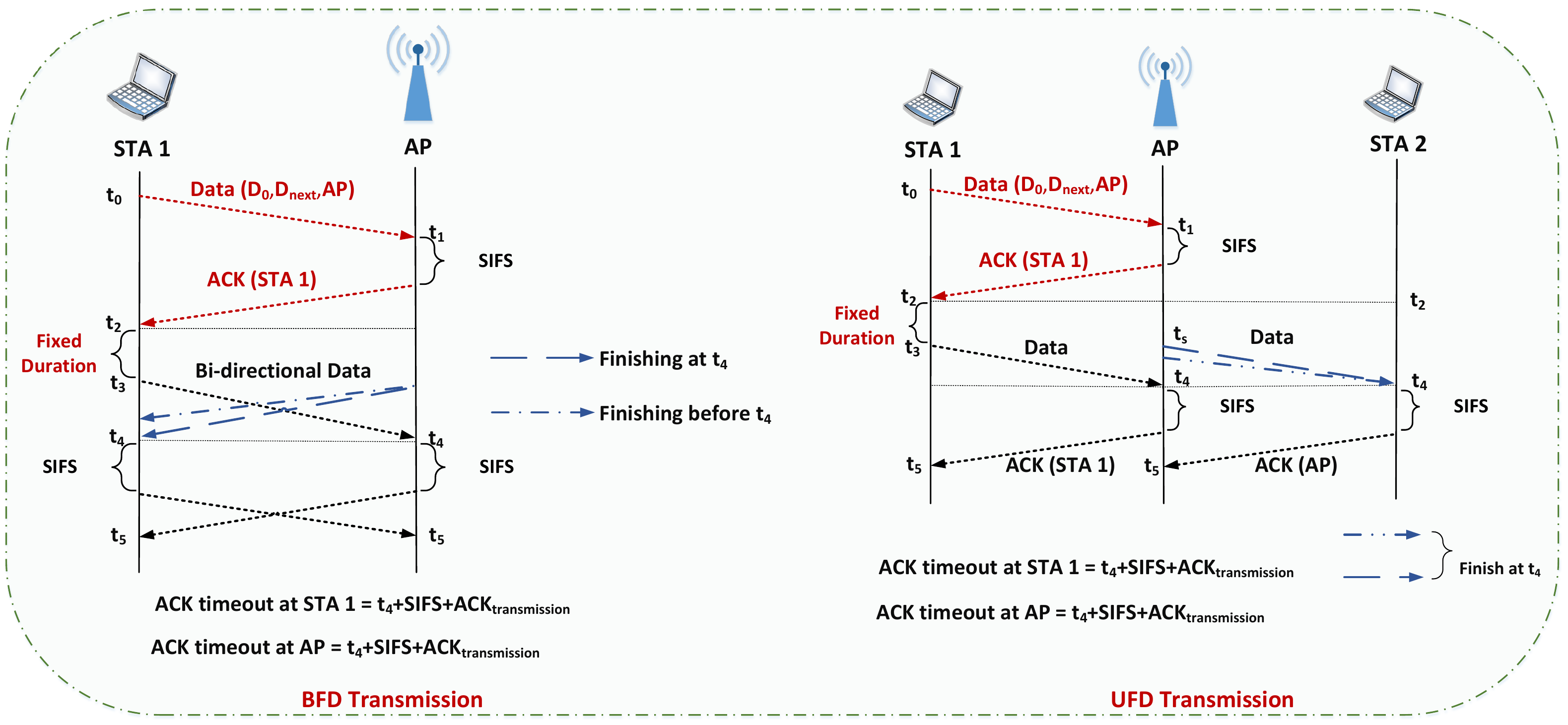}}
\caption{An illustration of the system model and the MAC layer operation of \textsf{MASTER} in (a). The timing aspects of  BFD and UFD transmissions are depicted in (b). \textcolor{black}{Please see \cite{aijaz_str_wcm} for further details with respect to timing aspects, particularly on ACK timeout setting.}}
\end{figure*}

\section{\textsf{MASTER} -- MAC Protocol Operation}
To describe the protocol operation of \textsf{MASTER}, we consider a single-cell, multi-user 802.11 WLAN wherein both FD and legacy HD STAs co-exist with either CSMA/CA or CSMA/ECA capabilities. We assume that the FD STAs employ necessary self-interference cancellation techniques at the PHY layer. Further, we assume that any FD or HD STA in the network is capable of adjusting its CST.  

\subsection{Capability Discovery}
Initially, nodes in the network should be able to discover FD and CSMA/ECA capabilities. \textcolor{black}{Similar to \cite{aijaz_str_wcm}, \textsf{MASTER} achieves capability discovery by embedding additional information in the \emph{Capability Information} (CI) field of the management frames.} This ensures backward compatibility for legacy nodes in the network. The CI field consists of 2 bytes, out of which 1 byte is reserved and is currently used to advertise different capabilities such as encryption, MIMO, quality-of-service (QoS) etc. The AP can periodically advertise its capabilities in the beacon frame. Hence, STAs in the network can learn if the AP is FD-capable and supports CSMA/ECA or not.
Both  FD and CSMA/ECA capabilities can be advertised through a change in any of
the reserved bits. Similarly, a STA can also inform the AP of its FD or CSMA/ECA capabilities via the association request frame. 

\subsection{Eligible Node Identification}
In case of UFD transmissions, it is particularly important that the two STAs which are simultaneously served by the AP are out of the interference range of each other, as otherwise the primary transmitter (STA 1 in Fig. \ref{fd_scenarios}) will interfere with the secondary receiver (STA 2 in Fig. \ref{fd_scenarios}). Therefore, the AP must know which nodes are eligible to become part of the UFD transmission. 
\textsf{MASTER} implements a standards compatible procedure for the AP to obtain this information. The IEEE 802.11k amendment \cite{802_11k}, which was introduced in 2008, supports radio resource measurement functionality for improving the performance of the network.  Most of the radio resource measurements can be exchanged in a request-report fashion. The requests and reports for radio resource measurements are sent in the
body of action frames. An action frame is a type of management frame that triggers an action. The IEEE 802.11k amendment defines the format and duration of radio resource measurements but does not specify when they have to be performed.

\emph{Frame measurement} is a specific type of radio resource measurement which is performed in a request-report fashion. In this report, the measuring STA could report the number of frames received, the power level, and the basic service set identifier (BSSID) for every transmitter address it listens to. In order to acquire the neighborhood information, the AP in \textsf{MASTER} can periodically send the ‘frame measurement’ request frame to each STA in the network. The respective STA can respond with a ‘frame measurement’ response frame. \textcolor{black}{The measurement response contains information about each neighbour of the STA sending the report, the associated power level and the BSSID. Based on the measurements received from different STAs, the AP can acquire the knowledge of eligible nodes for UFD
transmission. This is because each STA reports the neighboring STAs it can hear. By interpolating the overall information received from each STA, the AP can identify eligible nodes for UFD transmission. }It is worth noting that the IEEE 802.11k amendment has
been absorbed into the recent IEEE 802.11n-2016 standard.


\subsection{Establishing BFD and UFD Transmissions}
Next, we explain how CSMA/ECA can be exploited for establishing BFD and UFD transmissions. Consider the topology and the associated timeline shown in Fig. \ref{STR_txn1} where the circles represent the carrier sensing range of a node. STAs are able to carrier sense each other if their carrier sensing ranges overlap. Note that in WLANs, time is divided into three distinct types of timeslots: fixed length empty slots, collision slots, and successful slots. The collision and successful slots are much larger than empty slots. \textcolor{black}{On the horizontal axis, against each STA, the numerical value shows the number of slots left for the backoff to expire.} Consider that all the STAs support CSMA/ECA.  Since STA 1 and STA 3 have picked the same random backoff, their backoff counters are set to expire in the same timeslot. Hence, their transmissions will collide in the first instance. After collision, both STAs take a random backoff. In the second instance, the transmission attempt of STA 1 is successful. We assume that \(CW_{\text{min}}\) is set to \(10\) timeslots. Hence, STA 1 takes a deterministic backoff of \(4\) timeslots (see Section \ref{info_eca} for details). Assume that STA1 is FD-capable and the AP has data to transmit to it. Recall that the AP is aware of the FD and CSMA/ECA capabilities of STA 1. Based on the use of a deterministic backoff by STA 1, the AP has \emph{a priori} knowledge of when it will transmit again. Therefore, the use of CSMA/ECA provides the opportunity of a BFD transmission between AP and STA 1 through identification of potential opportunities for secondary transmission. \textcolor{black}{Note that the use of CSMA/ECA does not guarantee the occurrence of a BFD transmission. Any collision at the next transmission timeslot leads to the loss of STR opportunity.  }


\textcolor{black}{As discussed in \cite{aijaz_str_wcm}, it is particularly important that the secondary transmission (AP to STA 1) ends before or at the same time as the primary transmission (STA 1 to AP). This is mainly to resolve the ACK timeout issue\footnote{\textcolor{black}{The ACK timeout issue and the solution to overcome this issue has been discussed in detail in our previous work \cite{aijaz_str_wcm}.}}. By setting the ACK timeout duration (at both STA 1 and AP) to the sum of the transmission time for data, the short interframe spacing (SIFS) duration, and the transmission time of ACK, as shown in Fig. \ref{STR_txn2}, this issue can be avoided. In order to realize this ACK timeout, the aforementioned condition on finishing the secondary transmission before or at the same time as the primary transmission arises.} 
Note that STA 1 will specify the duration of the primary transmission in the Duration ID (henceforth referred to as the duration) field of the packet header. The significance of the duration field is that the STAs overhearing the transmission will set their network allocation vector (NAV) based on  this field. The NAV indicates for how long a STA must defer access to the medium. Using the duration information, the AP can simultaneously start sending a packet to STA 1. The transmission time of the secondary transmission depends on the payload and the modulation and coding scheme (MCS). Therefore, the AP must select the payload and MCS for the secondary transmission that ensures completion before or at the same time as the primary transmission. The secondary transmission can be further optimized as discussed later. 

Next, we explain the UFD transmission case. Consider the same topology as shown in Fig. \ref{STR_txn1}. As before, the transmissions of STA 1 and STA 3 collide in the first instance. In the second instance, the transmission of STA 1 is successful. Therefore, it takes a deterministic backoff of \(4\) timeslots. Hence, the AP knows when STA 1 will transmit again. However, unlike the previous case, we assume that the AP has data to send to STA 2 instead of STA 1. Therefore, the AP identifies an opportunity for engaging in a UFD transmission, based on the knowledge that STA 2 and STA 1 are out of the carrier sensing range of each other. As before, STA 1 will specify the duration of the primary transmission in the duration field of the packet header. Using this information, the AP can simultaneously start sending a packet to STA 2. The transmission time of the secondary transmission depends on the payload and the modulation and coding scheme (MCS). \textcolor{black}{To avoid the ACK timeout issue in the UFD case, it is particularly important that both the primary and the secondary transmissions finish at the same time \cite{aijaz_str_wcm}}. Therefore, the AP must select the payload and MCS for the secondary transmission that ensures completion at or before the primary transmission. As discussed in \cite{aijaz_str_wcm}, the AP can also adjust the start time of the secondary transmission to fulfil this constraint.
Timing aspect of both BFD and UFD transmissions is shown in Fig. \ref{STR_txn2}.

\begin{figure*}
\centering
\includegraphics[scale=0.28]{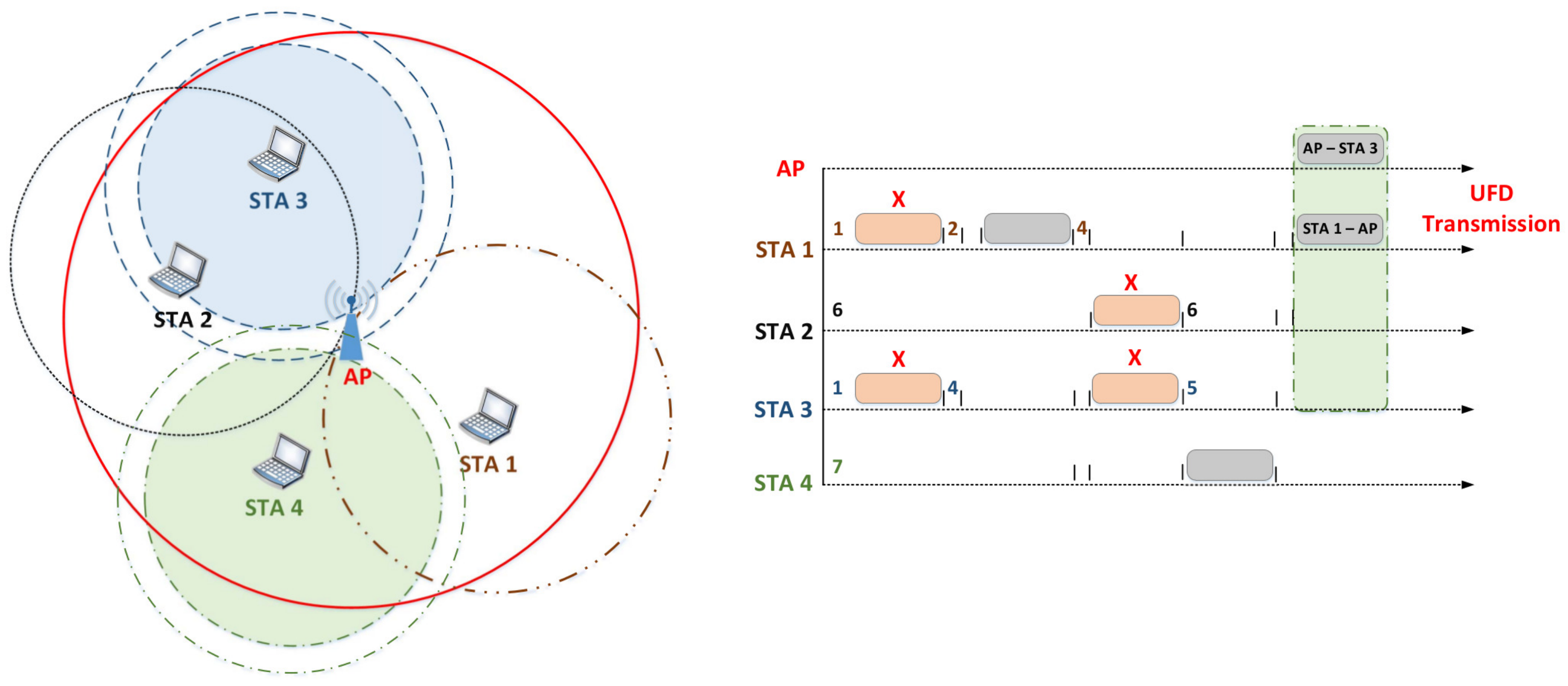}
\caption{An illustration of adaptive sensitivity control for creating UFD opportunities. Filled circle denotes the modified carrier sensing range of a STA. }
\label{STR_txn_AST}
\end{figure*}

\begin{table*}[]
\centering
\caption{Example of adaptive sensitivity control for the scenario depicted in Fig. \ref{STR_txn_AST}}
\label{cst_table}
\begin{tabular}{c|c|c|c|c|c}
\hline
\toprule
\begin{tabular}[c]{@{}c@{}}Node \\ ID\end{tabular} & \begin{tabular}[c]{@{}c@{}}RSSI of the AP\\ (A)\end{tabular} & \begin{tabular}[c]{@{}c@{}}Neighbor\\ ID\end{tabular} & \begin{tabular}[c]{@{}c@{}}RSSI of the Neighbor\\ (B)\end{tabular} & \begin{tabular}[c]{@{}c@{}}Tolerance\\ (C)\end{tabular} & \begin{tabular}[c]{@{}c@{}}Max. CST\\ \(\min\) (B+C, A)\end{tabular} \\ \hline
\midrule
\multirow{2}{*}{STA 1}                             & \multirow{2}{*}{-\(55\) dBm}                                     & STA 3                                                 & \(-77\) dBm                                                            & \(5\) dB                                                    &  \( \min (-77+5, -55) = -72\) dBm                                      \\ \cline{3-6} 
                                                   &                                                              & STA 4                                                 & \(-55\) dBm                                                            & \(5\) dB                                                    & \(\min (-55+5, -55) = -55\) dBm                                      \\ \hline
\toprule                                                   
\multirow{2}{*}{STA 2}                             & \multirow{2}{*}{-45 dBm}                                     & STA 3                                                 & \(-50\) dBm                                                            & \(5\) dB                                                    & \(\min (-50+5, -45) = -45\) dBm                                       \\ \cline{3-6} 
                                                   &                                                              & STA 4                                                 & \(-65\) dBm                                                            & \(5\) dB                                                    & \(\min (-65+5, -45) = -60\) dBm                                      \\ \hline
                                                   \toprule
\multirow{3}{*}{STA 3}                             & \multirow{3}{*}{-55 dBm}                                     & STA 1                                                 & \(-80\) dBm                                                            & \(5\) dB                                                    & \(\min (-80+5, -55) = -75\) dBm                                      \\ \cline{3-6} 
                                                   &                                                              & STA 2                                                 & \(-50\) dBm                                                            & \(5\) dB                                                    & \(\min (-50+5, -55) = -55\) dBm                                      \\ \cline{3-6} 
                                                   &                                                              & STA 4                                                 & \(-70\) dBm                                                            & \(5\) dB                                                    & \(\min (-70+5, -55) = -65\) dBm                                      \\ \hline
                                                   \toprule
\multirow{3}{*}{STA 4}                             & \multirow{3}{*}{-35 dBm}                                     & STA 1                                                 & \(-55\) dBm                                                            & \(5\) dB                                                    & \(\min (-55+5, -35) = -50\) dBm                                      \\ \cline{3-6} 
                                                   &                                                              & STA 2                                                 & \(-60\) dBm                                                            & \(5\) dB                                                    & \(\min (-60+5, -35) = -55\) dBm                                      \\ \cline{3-6} 
                                                   &                                                              & STA 3                                                 & \(-70\) dBm                                                            & \(5\) dB                                                    & \(\min (-70+5, -55) = -65\) dBm                                      \\ \hline
                                                   \toprule
\end{tabular}
\end{table*}

\subsection{Optimization of Secondary Transmission}
In general, the look up operation for selecting a suitable node and a suitable packet could be computationally expensive, potentially to the extent that it could lead to missing out the secondary transmission opportunity altogether. Hence, it is beneficial to have \emph{a priori} knowledge of this information. This is where there is merit in exploiting the CSMA/ECA capability further. When a CSMA/ECA-capable node (e.g., STA 1) sends a packet to its serving AP, assuming its transmission is likely to be successful, it includes the duration information for the next packet (referred to as \(D_{\text{next}}\)), in its buffer, in the header of the current packet. Note that the duration is dependent on the packet size and the chosen MCS. Therefore, by assuming that the MCS does not change for the packet that will immediately follow, STA 1 could
choose the next packet from the queue and, based on the known MCS information, calculate its duration. This information could be carried in any of the reserved fields of the packet header to maintain full backward compatibility i.e., the legacy nodes will simply ignore the content of reserved fields. It is worth noting that STA 1 could include
the payload size information instead of the duration information. However, in this case, the AP will have to assume that the next packet will arrive at the same MCS, and therefore, compute the duration. Thus, it does not matter whether duration or payload size is advertised by STA 1. However, to simplify implementation, it might be easier for STA 1 to advertise the duration information.

When the AP receives this packet (i.e., if the transmission of STA 1 is successful), it knows the time at which STA 1 is going to transmit again, owing to the use of a deterministic backoff. Therefore, it provides enough lead time to the AP to find an eligible node and a packet that can be served within time \(D_{\text{next}}\). When the first transmission completes successfully and the AP sends an ACK, STA 1 takes a deterministic back-off and starts the next transmission to the AP. \textcolor{black}{Since the AP knows the start time of this transmission, it selects the packet identified from the queue and
starts transmitting it to the secondary receiver at the same time as STA 1 is transmitting the packet to the AP.} There could be two possibilities: either the secondary transmission succeeds or it fails. In the former case, there is a clear benefit, in terms of throughput and medium utilization, over legacy HD approach. In the latter case, the performance would be the same as that of the legacy HD approach; hence, there is nothing to lose. 

Note that the signaling of the duration of the next transmission within the packet header of the first transmission by a CSMA/ECA capable node ensures maximum utilization of the secondary transmission opportunity.

\subsection{Creation of UFD Opportunities}
Next, we explain how adaptive sensitivity control can be exploited for achieving STR operation in 802.11 WLANs. Recall that a key requirement for  UFD transmission is that the two nodes simultaneously served by the AP are out of the carrier sensing range of each other. The AP can acquire the knowledge of eligible nodes for UFD transmissions based on the IEEE 802.11k measurement request-report procedure described earlier. However, such a solution only provides \emph{identification} of eligible nodes.

We propose a novel approach wherein nodes adapt their CSTs to turn a deaf ear to an ongoing transmission in anticipation for receiving a potential secondary transmission. Initially, the AP requests \emph{frame measurement} and \emph{link measurement} reports from all the STAs in the network. The former includes information about the received power level  for each neighbour of a node whereas the latter includes the same information for the link with the AP. Consider the topology shown in Fig. \ref{STR_txn_AST} and assume that all STAs support CSMA/ECA capability. After collecting the measurement reports from different STAs, the AP creates a table, which is exemplified as \tablename~\ref{cst_table}. Note that, based on the topology shown in Fig. \ref{STR_txn_AST}, the only eligible nodes for a UFD transmission are STA 1 and STA 2. It is desirable for the AP to have a list of potential target STAs that could be simultaneously served in a UFD transmission. Therefore, the AP temporarily creates eligible nodes to maximize UFD opportunities as explained below.

Let, STA 1 be the primary transmitter. We observe from \tablename~\ref{cst_table} that the STA 3 can hear STA 1 at \(-80\) dBm. Thus, if the AP were to choose STA 3 as a potential target for a secondary transmission, STA 3 could hear both the primary transmission from STA 1 as well as the secondary transmission from the AP, which would lead to a collision at STA 3. If STA 3 adapts its CST such that it  turns a deaf ear to the transmission from STA 1 but not for the transmission from the AP, then both transmissions could go in parallel and potentially succeed.	This can be accomplished as follows. Note that the carrier sensing range of a node is inversely proportional to its CST. From \tablename~\ref{cst_table} we observe that after adding a fixed margin to the \textcolor{black}{received signal strength indicator (RSSI)}  of STA 1, the CST at STA 3 is still less than the RSSI of the AP. Hence, STA 3 will still be able to hear the AP. Similar situation arises at STA 4 as well. A key question that arises from an implementation perspective is that how a node decides about updating its CST and for how long? We assume that the AP advertises the CSMA/ECA capabilities of different STAs via the beacon frame. \textcolor{black}{Therefore, all the STAs which are capable of CSMA/ECA are aware of the next transmission attempt from STA 1}. Hence, these STAs will update their CST, as shown in \tablename~\ref{cst_table}, when the primary transmission starts. Moreover, the CST adaptation would last for the duration of the primary transmission. Hence, in context of the example in question, both STA 3 and STA 4 will update their CST. Once the primary transmission completes, all STAs revert the CST to its default value. Note that all of these STAs will wait in anticipation of a secondary transmission. The AP has the knowledge of which STAs could be effective targets, and therefore, it will select one of these STAs for the secondary transmission. In the worst case, even if none of these STAs is chosen (e.g. AP ends up choosing STA 2 which is a natural eligible STA without
even tweaking its CST), they can simply revert back to their default CST at the end of the primary transmission. This approach simplifies implementation and enables
an opportunistic solution wherein secondary targets are ready irrespective
of whether they are chosen or not.

Next, we explain how a node may not be eligible for a secondary transmission after adapting its CST. Assume that STA 4 is the primary transmitter. According to \tablename~\ref{cst_table}, STA 1 can hear STA 4 at \(-55\) dBm. Thus, if STA 1 adapts its CST (\(-55+5=-50\) dBm), it will drop below the RSSI from the AP, and therefore, a secondary transmission involving STA 1 is unlikely to succeed as it will not be able to hear the AP in the first place. Thus, the AP will exclude STA 1 from its list of secondary targets when
STA 4 is the primary transmitter. This illustrates the simplicity and practicality of the proposed solution. \textcolor{black}{Note that the tolerance margin is a fixed parameter (for all STAs) and must be selected to account for the variations in channel conditions.} \qed

\textcolor{black}{\emph{Remark 1} -- The proposed algorithmic framework to identify eligible nodes and to create opportunities for UFD transmissions relies on IEEE 802.11k-based measurement reports from STAs in the network. Despite its simplicity, it may generate some overhead which is dependent on the frequency of such reports. However, the benefits of STR operation largely outweigh the overhead of realizing this functionality.} 


\begin{figure*}
\label{combined_str_gain}
\centering
\subfloat[]{\label{STR_Gain}\includegraphics[scale=0.22]{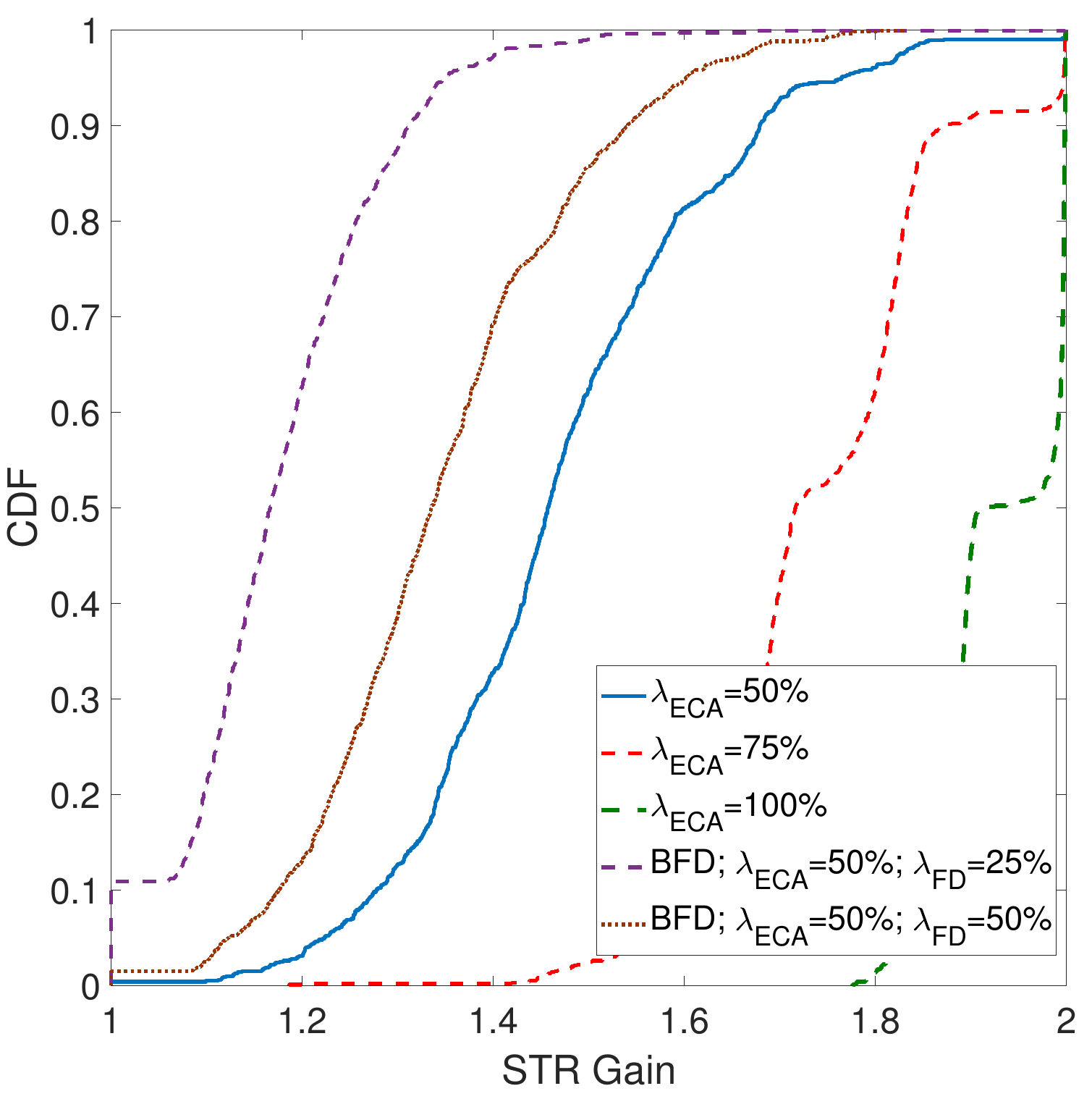}} \
\subfloat[]{\label{STR_Gain_UFD}\includegraphics[scale=0.22]{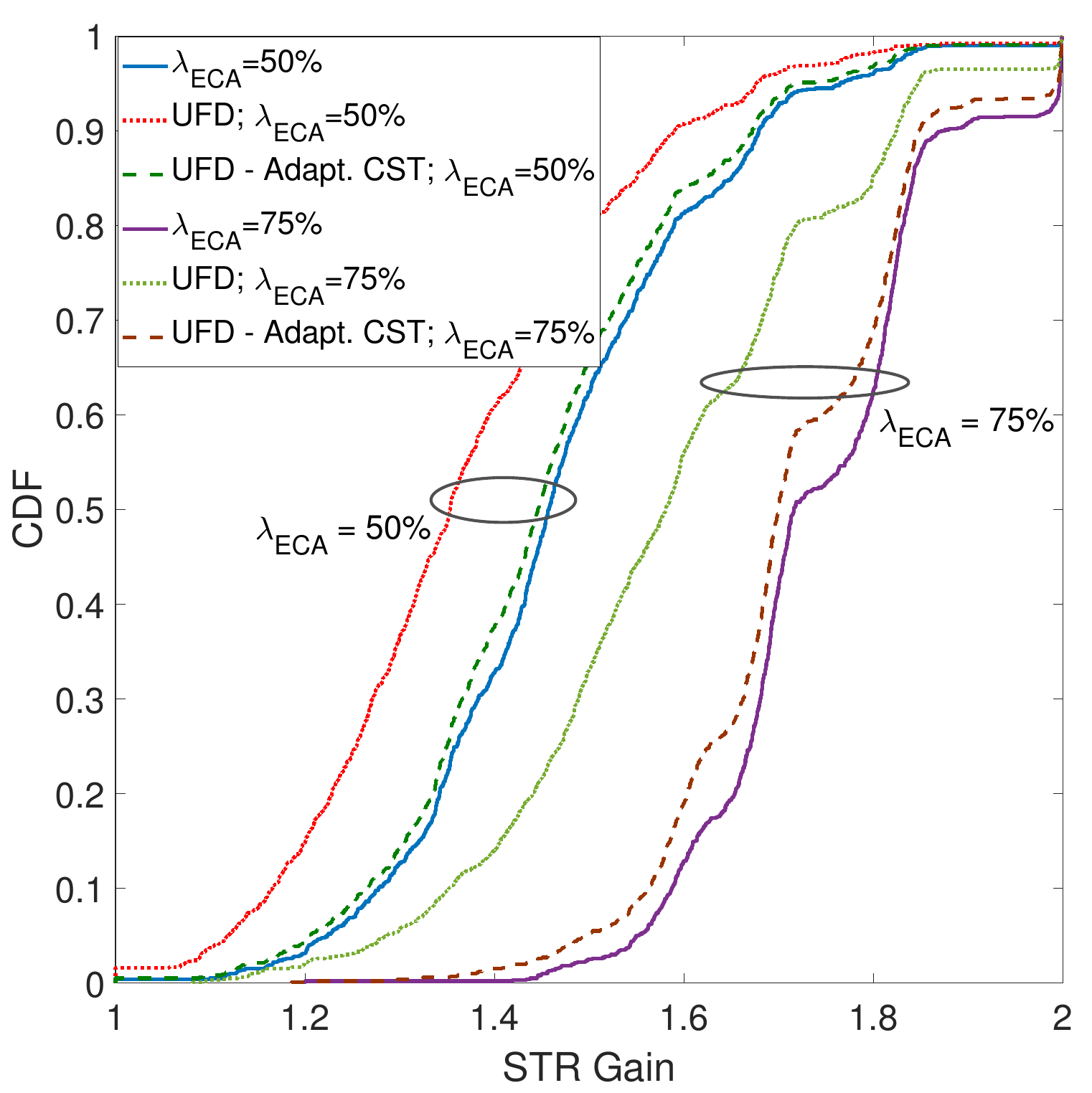}} \
\subfloat[]{\label{STR_Gain_U_rad}\includegraphics[scale=0.22]{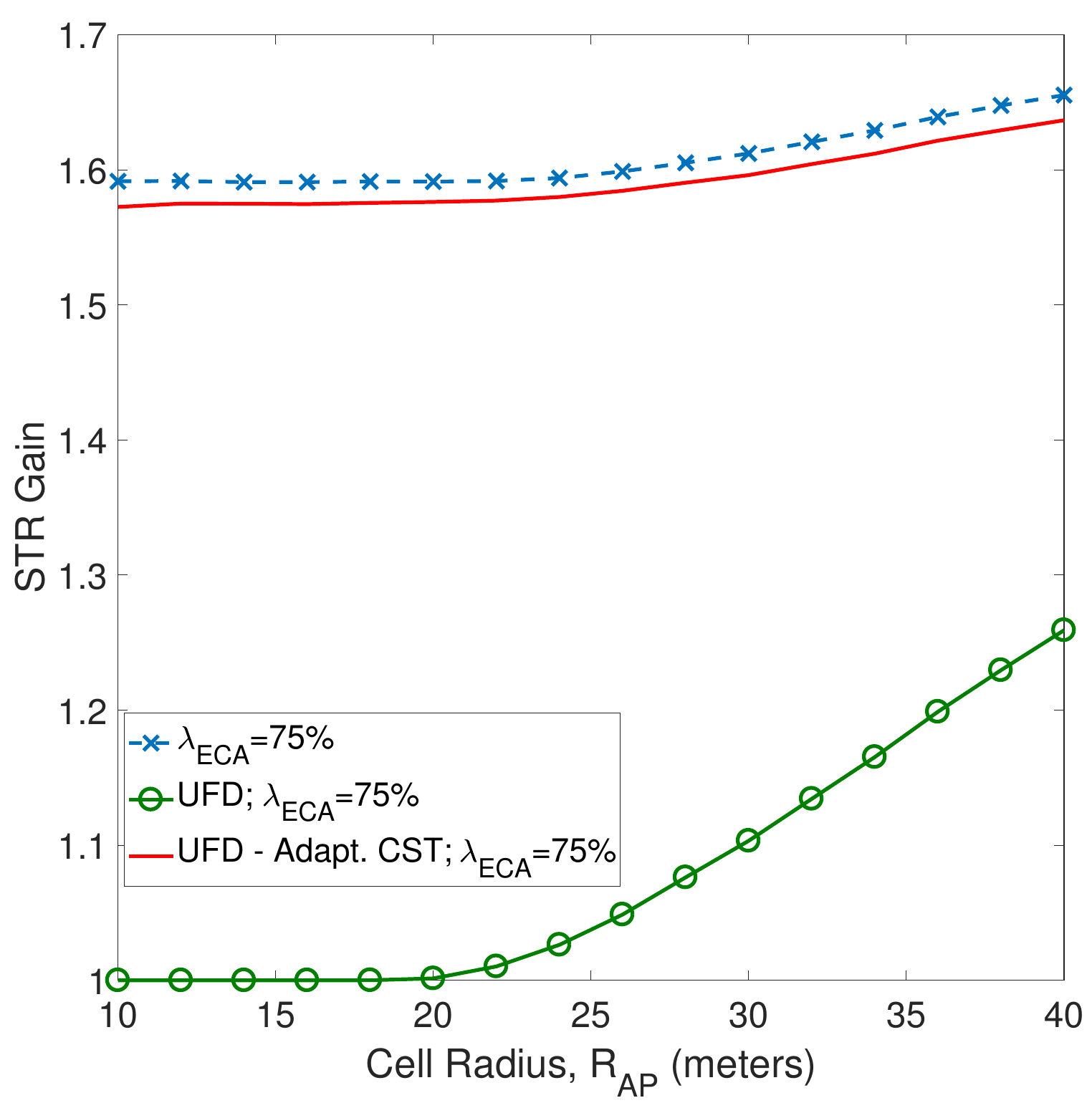}} \qquad
\subfloat[]{\label{UFD_opps}\includegraphics[scale=0.22]{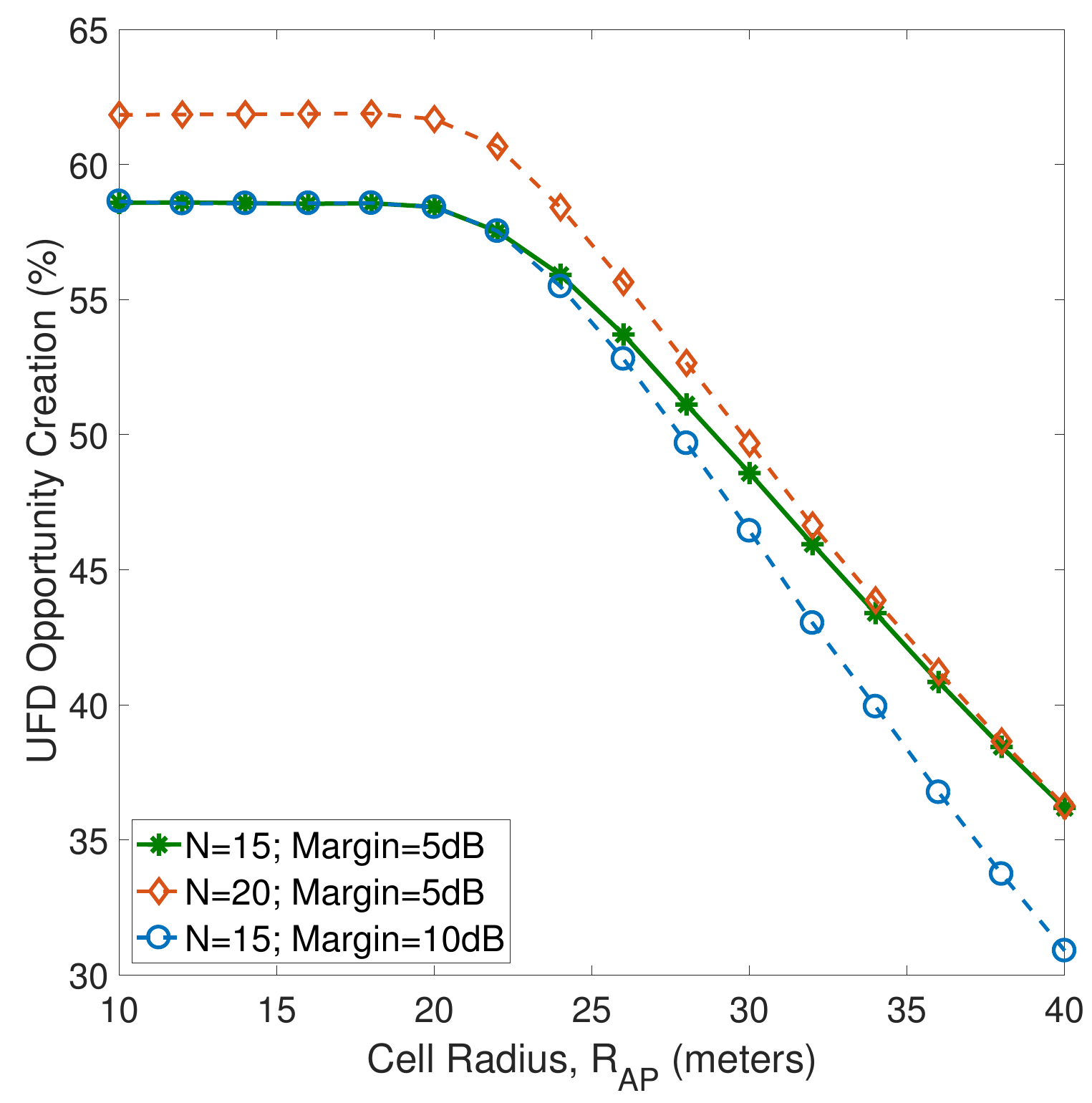}} \
\subfloat[]{\label{G_SNR}\includegraphics[scale=0.22]{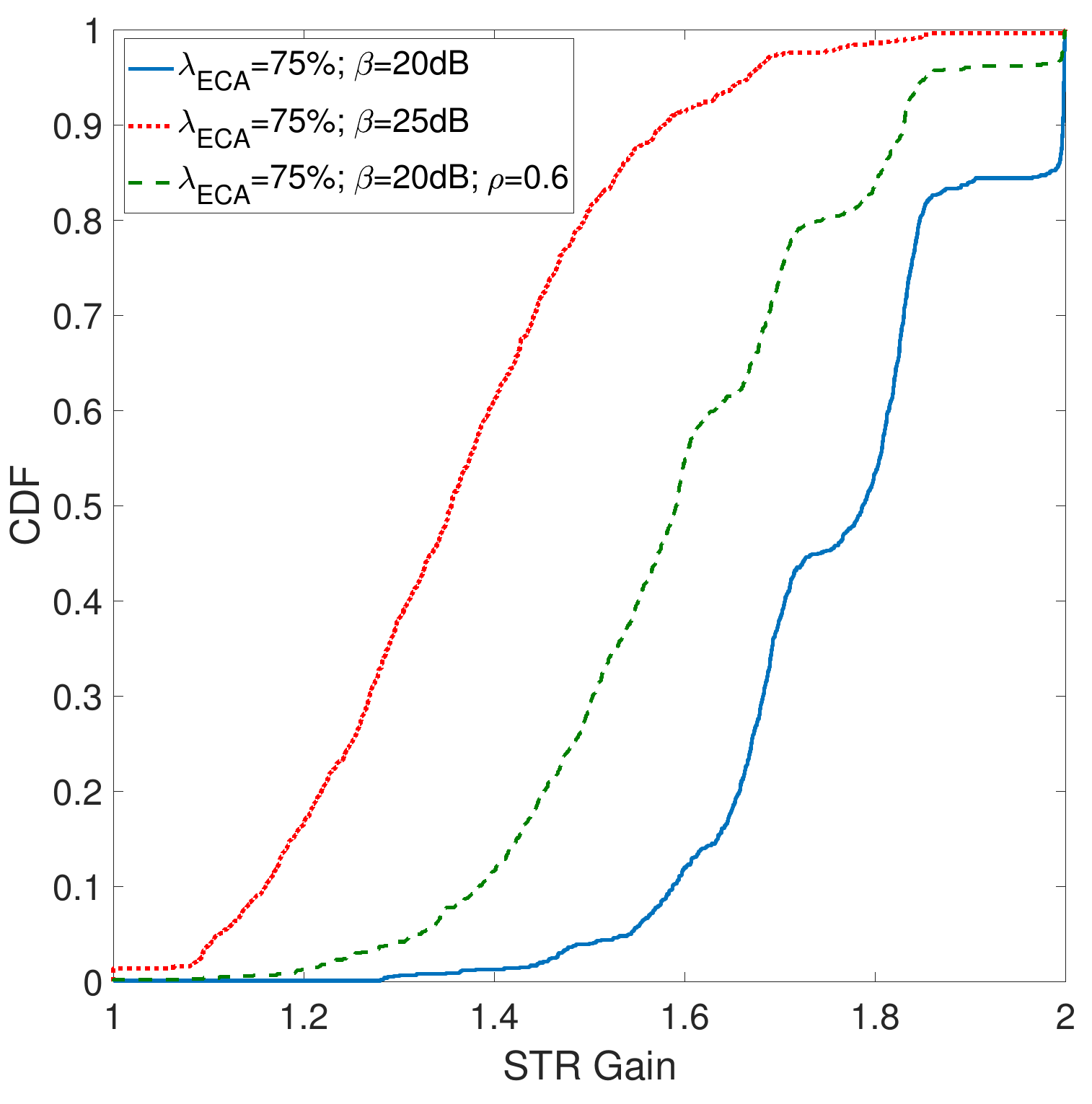}} \
\subfloat[]{\label{G_CW}\includegraphics[scale=0.22]{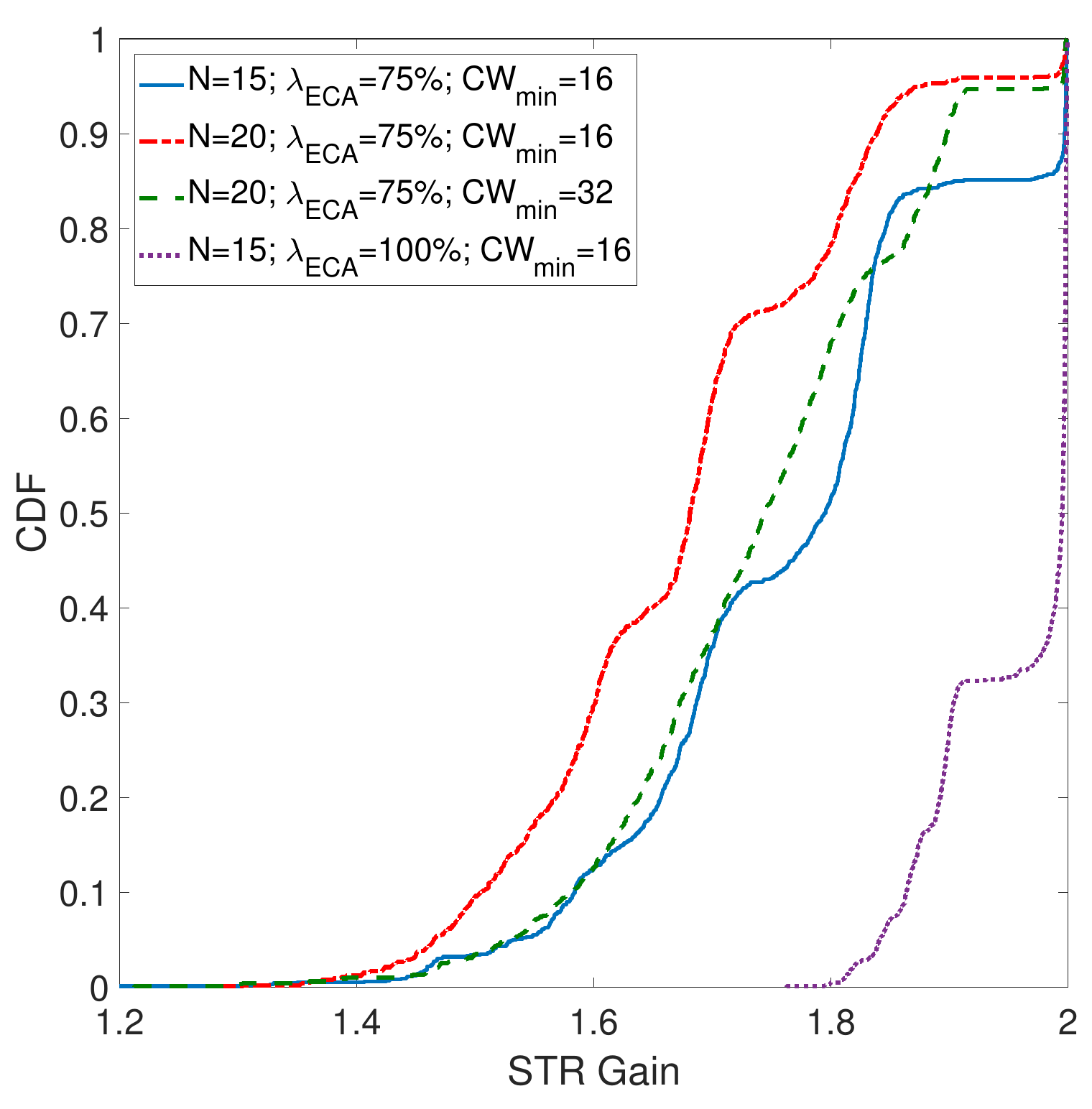}} 
\caption{Performance evaluation of \textsf{MASTER} (a):  CDF of STR Gain for CSMA/ECA and BFD case (\(N=15, R_{AP}=35 \ \text{m}, \beta=20 \ \text{dB}\)); (b):  CDF of STR Gain for CSMA/ECA and UFD case (\(N=15, R_{AP}=35 \ \text{m}, \beta=20 \ \text{dB}) \); STR Gain against cell radius (\(N=15, \beta=20 \ \text{dB}\)); (d) UFD opportunity creation against cell radius; (e) CDF of STR Gain (\(N=15, R_{AP}=35 \ \text{m}\)); (f) CDF of STR Gain (\(R_{AP}=35 \ \text{m}, \beta=20 \ \text{dB} \)). }
\end{figure*}

\section{Performance Evaluation}\label{sect_perf}
We evaluate the performance of \textsf{MASTER} through system-level simulations. Our customized simulator implements IEEE 802.11 DCF specifications with both CSMA/CA and CSMA/ECA and the proposed protocol for enabling STR mode. We adopt the \(19\)-cell hexagonal grid model such that \(N\) STAs are uniformly distributed in the coverage of each AP with a radius of \(R_{AP}\). The channel model accounts for large-scale path loss and small-scale Rayleigh fading. The transmit power for both the APs and the STAs is set to \(14\) dBm. We consider a co-existence scenario wherein the percentage of CSMA/ECA and legacy CSMA/CA  STAs is given by  \(\lambda_{ECA}\) and \(\lambda_{CA}\), respectively. Further, the percentage of FD and legacy HD  STAs is given by \(\lambda_{FD}\) and \(\lambda_{HD}\), respectively. The default CST is set to \(-82\) dBm. Finally, we adopt an experimentally characterized model \cite{exp_FD} for residual self-interference (RSI) in FD nodes. The traffic model is assumed to be backlogged with packet payload of \(1000\) bytes, MAC header of \(272\) bits, and PHY header of \(128\) bits. The transmission rate is set to \(54\) Mbps.  We perform Monte Carlo simulations on different user distributions. 

We define the STR gain as \(\theta_{STR}=\chi_{STR}/\chi_{L}\), where \(\chi_{STR}\) and \(\chi_{L}\) denote the average throughput of the network with STR and legacy mode of operation, respectively. Fig. \ref{STR_Gain} plots the cumulative distribution function (CDF) of the gain of STR for CSMA/ECA in different scenarios. Note that the STR gain increases as \(\lambda_{ECA}\) increases. This is due to the fact that a larger number of CSMA/ECA-capable STAs increases the possibilities of FD transmissions in the network. In \(80\%\) of the scenarios, STR gain of up to \(1.62\) can be achieved when \(50\%\) of the STAs are CSMA/ECA-capable. A gain of up to \(1.8\) and nearly \(2\) can be acheived when \(75\%\) and \(100\%\) STAs in the network are CSMA/ECA-capable, respectively. Further, the gain for BFD case increases as \(\lambda_{FD}\) increases. This is due to higher number of BFD transmissions in the network. In \(80\%\) of the scenarios, a gain of \(1.24\) is achieved by \(25\%\) FD-capable STAs in the network.

\begin{table*}
		\caption{Qualitative Comparison of Different MAC Protocols}
		\begin{center}
			\begin{tabular}{lcccccc}
				\hline	
				\toprule
		\textbf{Feature / Protocol} &  \textbf{A-Duplex} \cite{a-duplex} & \textbf{FD-MAC} \cite{fd_mac} & \textbf{PoC-MAC} \cite{PoCMAC} & \textbf{S-CW FD MAC} \cite{scw-fd} & \textbf{\textsf{STR-MAC}} \cite{aijaz_str_wcm} & \textbf{\textsf{MASTER}}  \\\hline
				\midrule			
				Handshake-free Operation  & No & No & No & No & No & Yes \\
				FD/HD Co-existence  & No & Yes & No & Yes & Yes & Yes \\
				BFD Transmissions & No & Yes & No & Yes & Yes & Yes \\
				UFD Transmissions  & Yes & No & Yes & No & Yes & Yes \\
				UFD Opportunity Creation  & No & No & No & No & No & Yes \\
				\hline
			\end{tabular}
		\end{center}
			\label{qual_comp}
	\end{table*}

Fig. \ref{STR_Gain_UFD} shows the CDF of STR gain for CSMA/ECA and UFD scenario. Note that with natural UFD opportunities the STR gain of UFD is significantly lower than the maximum possible gain provided by CSMA/ECA. However, with creation of UFD opportunities, through adaptation of the CST, the maximum achievable gain can be approached. For instance, with \(75\%\) CSMA/ECA-capable STAs in the network, a gain of up to \(1.8\) is achievable in \(80\%\) of the times. With natural UFD transmissions, the STR gain of UFD is \(1.7\), whereas with CST adaptation, the upper bound of \(1.8\) can be achieved. The impact of CST adaptation is much more significant in dense deployments, i.e., when the cell radii are smaller. For instance, in Fig. \ref{STR_Gain_U_rad}, the STR gain of UFD is nearly \(1\) for cell radii of up to \(20\) meters, i.e., no gain is achievable as no natural eligible nodes exist. However, through CST adaptation, the upper bound of STR gain, i.e., \(1.6\) can be approached. The results demonstrate that CST adaptation is particularly attractive for creation of UFD opportunities in dense deployments, in order to exploit the gains provided by CSMA/ECA. The creation of UFD opportunities is also dependent on the margin used for CST adaptation. As shown by the results in Fig. \ref{UFD_opps}, for a fixed number of STAs in the network, increasing the margin value from \(5\) dB to \(10\) dB results in reduction of UFD opportunities that can be created. This is because aggressive CST adaptation potentially leads to the scenario where the CST drops below the RSSI from the AP. Such scenarios are more likely arise when cell radii are larger. For example, when the cell radius is \(35\) meters, UFD opportunity creation drops by nearly \(15\%\) by increasing the margin to \(10\) dB. 

Fig. \ref{G_SNR} evaluates the impact of link-level signal-to-interference-plus-noise-ratio (SINR) on STR gain. The gain of STR deceases as the SINR threshold, \(\beta\) increases. This is because the  probability of a successful transmission reduces with stringent SINR requirements. The STR gain is also dependent on the RSI. A lower self-interference cancellation capability (\(\rho = 0.6\)) reduces the STR gain due to higher RSI.

Fig. \ref{G_CW} evaluates the impact of different parameters on the STR gain provided by CSMA/ECA. As shown by the results, the gain is dependent on the number of CSMA/ECA-capable STAs in the network and the minimum contention window, \(CW_{\text{min}}\). For a fixed \(CW_{\text{min}}\), increasing the number of STAs results in reduction of the STR gain. This is because of more collisions in the network due to which the throughput of CSMA/ECA decreases. One solution to increase the throughput of CSMA/ECA with larger number of STAs is to increase \(CW_{\text{min}}\). For instance, when \(N=20\), increasing \(CW_{\text{min}}\) to \(32\) yields nearly similar STR gain as \(N=15\) with \(CW_{\text{min}}\) of \(16\).

Finally, a qualitative comparison of \textsf{MASTER} against the prominent state-of-the-art protocols is given in \tablename~\ref{qual_comp}.

\section{Concluding Remarks}\label{sect_cr}
This paper proposed \textsf{MASTER} which provides a novel solution for enabling STR operation in 802.11 WLANs through CSMA/ECA and adaptive sensitivity control techniques. It exploits the key capability of CSMA/ECA for achieving STR operation without the need for any handshake mechanism. \textsf{MASTER} not only supports the co-existence of FD and legacy HD STAs but also of CSMA/ECA and legacy CSMA/CA nodes.  \textsf{MASTER} exploits adaptation of CST for creation of UFD opportunities.  Performance evaluation demonstrates that CSMA/ECA provides significant STR opportunities that can be exploited by BFD or UFD transmissions. Moreover, the creation of UFD opportunities through CST adaptation is crucial in achieving the gains provided by CSMA/ECA, particularly in dense deployments. 







\bibliographystyle{IEEEtran}

\bibliography{IEEEabrv,mybibfile}

\begin{thebibliography}{10}
\providecommand{\url}[1]{#1}
\csname url@samestyle\endcsname
\providecommand{\newblock}{\relax}
\providecommand{\bibinfo}[2]{#2}
\providecommand{\BIBentrySTDinterwordspacing}{\spaceskip=0pt\relax}
\providecommand{\BIBentryALTinterwordstretchfactor}{4}
\providecommand{\BIBentryALTinterwordspacing}{\spaceskip=\fontdimen2\font plus
\BIBentryALTinterwordstretchfactor\fontdimen3\font minus
  \fontdimen4\font\relax}
\providecommand{\BIBforeignlanguage}[2]{{%
\expandafter\ifx\csname l@#1\endcsname\relax
\typeout{** WARNING: IEEEtran.bst: No hyphenation pattern has been}%
\typeout{** loaded for the language `#1'. Using the pattern for}%
\typeout{** the default language instead.}%
\else
\language=\csname l@#1\endcsname
\fi
#2}}
\providecommand{\BIBdecl}{\relax}
\BIBdecl

\bibitem{survey_hew}
H.~A. Omar \emph{et~al.}, ``{A Survey on High Efficiency Wireless Local Area
  Networks: Next Generation WiFi},'' \emph{IEEE Commun. Surveys Tuts.},
  vol.~18, no.~4, pp. 2315--2344, Fourthquarter 2016.

\bibitem{FD_SIC}
A.~Sabharwal \emph{et~al.}, ``{In-Band Full-Duplex Wireless: Challenges and
  Opportunities},'' \emph{IEEE J. Sel. Areas Commun.}, vol.~32, no.~9, pp.
  1637--1652, Sept 2014.

\bibitem{FD_Phy_Mac}
D.~Kim, H.~Lee, and D.~Hong, ``{A Survey of In-Band Full-Duplex Transmission:
  From the Perspective of PHY and MAC Layers},'' \emph{IEEE Commun. Surveys
  Tuts.}, vol.~17, no.~4, pp. 2017--2046, Fourthquarter 2015.

\bibitem{xfdr_wcm}
\BIBentryALTinterwordspacing
M.~O. Al-Kadri, A.~Aijaz, and A.~Nallanathan, ``{X-FDR: A Cross-Layer Routing
  Protocol for Multi-hop Full-Duplex Wireless Networks},'' \emph{IEEE Wireless
  Commun.}, 2018. [Online]. Available: \url{https://arxiv.org/abs/1806.01737}
\BIBentrySTDinterwordspacing

\bibitem{aijaz_str_wcm}
\BIBentryALTinterwordspacing
A.~Aijaz and P.~Kulkarni, ``{Simultaneous Transmit and Receive Operation in
  Next Generation IEEE 802.11 WLANs: A MAC Protocol Design Approach},''
  \emph{IEEE Wireless Commun.}, 2017. [Online]. Available:
  \url{https://arxiv.org/abs/1706.07544}
\BIBentrySTDinterwordspacing

\bibitem{DCF}
``{Information technology--Telecommunications and information exchange between
  systems Local and metropolitan area networks--Specific requirements Part 11:
  Wireless LAN Medium Access Control (MAC) and Physical Layer (PHY)
  Specifications},'' \emph{ISO/IEC/IEEE 8802-11:2012(E) (Revison of
  ISO/IEC/IEEE 8802-11-2005 and Amendments)}, pp. 1--2798, Nov 2012.

\bibitem{CSMA_ECA_Perf1}
L.~Sanabria-Russo, J.~Barcelo, B.~Bellalta, and F.~Gringoli, ``{A High
  Efficiency MAC Protocol for WLANs: Providing Fairness in Dense Scenarios},''
  \emph{IEEE/ACM Trans. on Netw.}, vol.~25, no.~1, pp. 492--505, Feb 2017.

\bibitem{aijaz_DSC}
A.~Aijaz and P.~Kulkarni, ``{On Performance Evaluation of Dynamic Sensitivity
  Control Techniques in Next-Generation WLANs},'' \emph{IEEE J. Syst.},
  vol.~PP, no.~99, pp. 1--4, 2018.

\bibitem{ACS_Survey}
C.~Thorpe and L.~Murphy, ``{A Survey of Adaptive Carrier Sensing Mechanisms for
  IEEE 802.11 Wireless Networks},'' \emph{IEEE Commun. Surveys Tuts.}, vol.~16,
  no.~3, pp. 1266--1293, Third 2014.

\bibitem{a-duplex}
A.~Tang and X.~Wang, ``{A-Duplex: Medium Access Control for Efficient
  Coexistence Between Full-Duplex and Half-Duplex Communications},'' \emph{IEEE
  Trans. Wireless Commun.}, vol.~14, no.~10, pp. 5871--5885, Oct 2015.

\bibitem{fd_mac}
M.~Duarte \emph{et~al.}, ``{Design and Characterization of a Full-Duplex
  Multiantenna System for WiFi Networks},'' \emph{IEEE Trans. Veh. Technol.},
  vol.~63, no.~3, pp. 1160--1177, March 2014.

\bibitem{PoCMAC}
W.~Choi, H.~Lim, and A.~Sabharwal, ``{Power-Controlled Medium Access Control
  Protocol for Full-Duplex WiFi Networks},'' \emph{IEEE Trans. Wireless
  Commun.}, vol.~14, no.~7, pp. 3601--3613, July 2015.

\bibitem{scw-fd}
D.~Marlali and O.~Gurbuz, ``{S-CW FD: A MAC Protocol for Full-Duplex in
  Wireless Local Area Networks},'' in \emph{IEEE Wireless Communications and
  Networking Conference (WCNC)}, April 2016, pp. 1--6.

\bibitem{FD_WCL}
M.~O. Al-Kadri, A.~Aijaz, and A.~Nallanathan, ``{An Energy-Efficient
  Full-Duplex MAC Protocol for Distributed Wireless Networks},'' \emph{IEEE
  Wireless Commun. Lett.}, vol.~5, no.~1, pp. 44--47, Feb 2016.

\bibitem{802_11k}
``{IEEE Standard for Information technology-- Local and metropolitan area
  networks-- Specific requirements-- Part 11: Wireless LAN Medium Access
  Control (MAC)and Physical Layer (PHY) Specifications Amendment 1: Radio
  Resource Measurement of Wireless LANs},'' \emph{IEEE Std 802.11k-2008
  (Amendment to IEEE Std 802.11-2007)}, pp. 1--244, June 2008.

\bibitem{exp_FD}
M.~Duarte, C.~Dick, and A.~Sabharwal, ``{Experiment-Driven Characterization of
  Full-Duplex Wireless Systems},'' \emph{IEEE Trans. Wireless Commun.},
  vol.~11, no.~12, pp. 4296--4307, Dec 2012.

\end{thebibliography}
%

\end{document}